\documentclass[newversion]{aa}
\usepackage{graphicx}

\begin{document}


   \title{The Infrared Ca\,II lines in Sunspot Umbrae}

   \author{W. Kollatschny\inst{1}
          \and G. Stellmacher\inst{2}
          \and E. Wiehr\inst{1} 
          \and M.\,A. Fallipou\inst{2}}

   \offprints{E. Wiehr}

   \mail{ewiehr@astrophysik.uni-goettingen.de}

   \institute{Institut f\" ur Astrophysik der Universit\"at,
              Friedrich-Hund-Platz 1, 37077 G\"ottingen, Germany
              \and
              Institute d'Astrophysique (IAP), 98 bis Blvd. d'Arago, 
              75014 Paris, France}

   \date{Received Aug. 8, 1979; accepted Aug. 30, 1979}

\abstract
{}
{We present an empirical working model for sunspot umbrae which equally describes 
observed continuum intensities and line profiles.}
{The wings of the infrared Ca\,II lines depend sensitively on the temperature 
gradient at $-0.6<log \tau_{0.5}<+0.3$ but not essentially on the absolute  
value of $T$. These lines are observed to remain almost unchanged from photosphere to 
umbra and are thus insensitive to parasitic light. It is also shown that the infrared 
K\,I\,7699\AA{} line is suitable for umbral spectroscopy since it is not seriously 
blended, its continuum is well defined and it is less influenced by parasitic light 
as compared to lines in the visible spectrum, due to the smaller umbal contrast.}
{Calculations show that the umbral gradient $dT/d\tau$, required to fit the Ca\,II 
triplet lines, strongly conflicts with the observed profiles of K\,I\,7699, Na\,D$_2$ 
and Fe\,I\,5434 (g=0), even when assuming vanishing Fe\,II lines for a maximum correction 
of parasitic light. It is shown that the discrepancy from the different line profiles 
may be removed by adopting an opacity enhancement as introduced by Zwaan (1974) from 
a discussion of continuum contrasts alone. The finally proposed umbral working model 
is very close to a scaled model of the quiet Sun with $T_{eff}= 4000$K thus resembling
a M0 rather than a K5 stellar atmosphere.} 
{}

\maketitle

%
%

\section{Introduction}

A sensitive method for the determination of the empirical model atmospheric of a
sunspot umbra is the investigating line profiles. The infrared Ca\,II triplet lines 
(at 8498, 8542, and 8662\AA{}) have so far not been considered. Calculations show that 
the wings of these lines are very sensitive to the temperature {\it gradient} at 
$-0.6<log\tau_{0.5}<+0.3$ but much less sensitive to the absolute value of the 
temperature. Cursory inspection of umbral spectra shows that these triplet lines 
are largely unchanged when compared to the spectrum of the neighboring photosphere. 
This implies an almost negligible influence of parasitic light, which anyway is 
reduced with respect to the visible region due to the smaller continuum contrast. 
In addition, the wavelength region around 8500\AA{} seems not seriously perturbed 
by purely umbral lines.

\section{Observations}

We investigated the infrared Ca\,II lines from several sunspot spectra observed in 
summer 1967, 1968, and 1979 at the Locarno solar station of the G\"ottingen observatory 
with the plane grating spectrograph (Czerny-Turner type) at high dispersion (0.2\AA/mm) 
and also with the concave grating spectrograph (Pasche-Rungen mounting) at low 
dispersion (2.5\AA/mm). Microphotometry of the infrared Ca\,II lines in these umbral 
spectra consistently shows no significant difference with the simultaneously observed 
line profiles in the neighboring photosphere. For further confirmation of this interesting 
result we recorded photoelectrically the line-to-continuum ratio of these Ca\,II lines 
in a spot near the disc center on July 7, 1979, and obtain the same result.

Those of our observed Ca\,II IR lines which occasionally show emission cores, always 
yield symmetric ones for Ca\,II\,8542 but asymmetric ones for 8Ca\,II\,498 and 
Ca\,II\,8662 with opposite sign: Ca\,II\,8498 has a steeper blue, Ca\,II\,8662 a 
steeper red flank of its core emission. 

In addition to these lines, spectra of the strong K\,I\,7699 line were investigated in 
the same manner. In contrast to the Ca\,II lines, this K\,I line is not unchanged and 
therefore strongly affected by parasitic light. A maximum correction has been applied 
by assuming Fe\,II\,7711 to vanish in umbrae (cf. Stellmacher and Wiehr, 1970). A first 
photoelectric recording of this K\,I line was obtained in the July 7, 1979, spot with 
particular care for parasitic light correction, which typically consists of 5\% penumbral 
blurring and $<1$\% photospheric scattering (in units of the photospheric continuum 
intensity).

   \begin{figure*}[ht]     
   \hspace{0mm}\includegraphics[width=18.2cm]{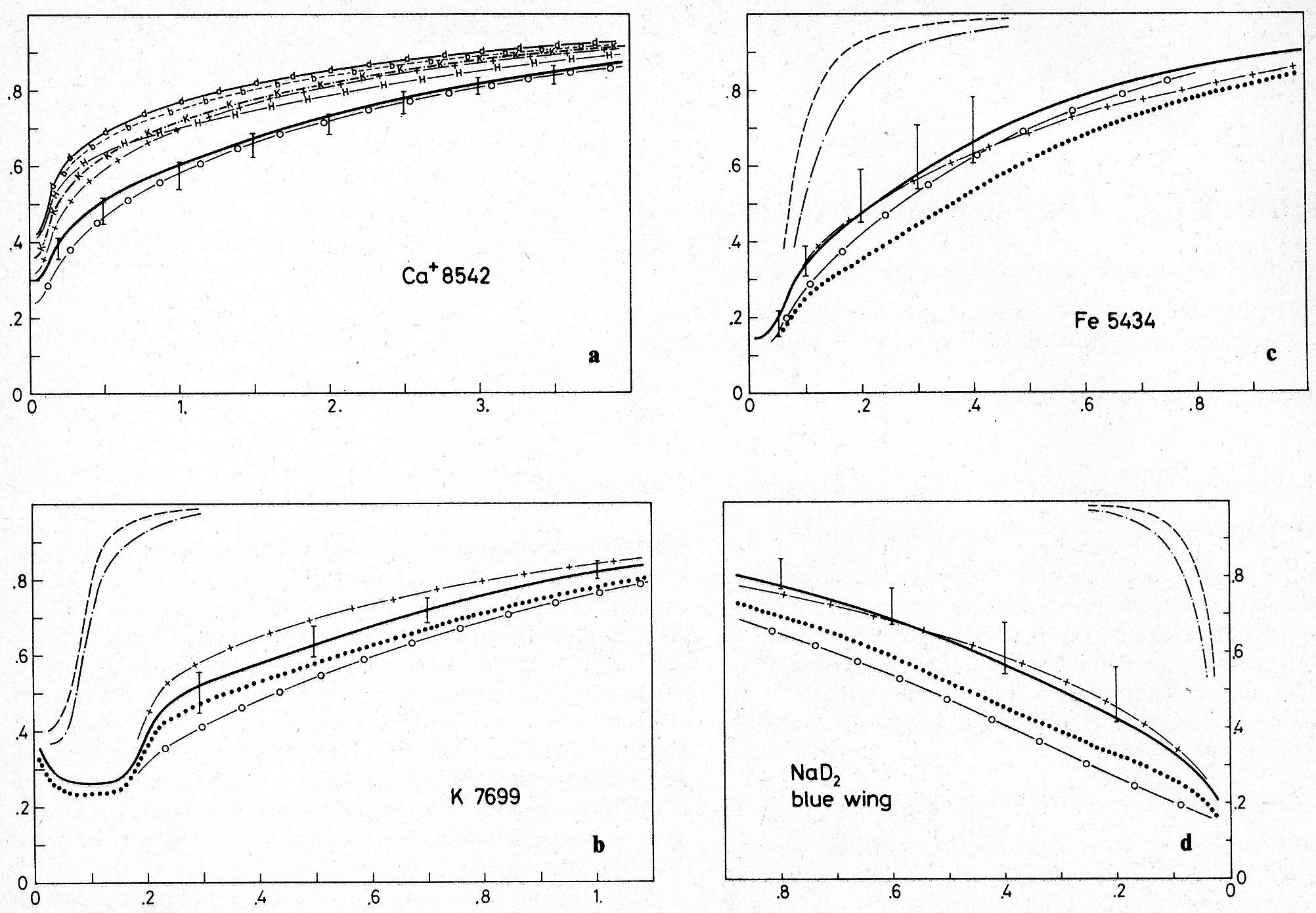}
   \caption{Observed lines profiles in sunspots: vertical bars give raw and maximum 
straylight-corrected values. Comparison with calculated profiles using the models 
by Kjeldseth-Moe \& Maltby dark (d) and bright (b), Stellmacher \& Wiehr (+), 
Henoux (H), Kneer (K), Zwaan with $q(\lambda)=1.05$ (O), and model M4 of this paper 
with q=1.0 ({\it full line}). The four panels give: {\bf [a]} Ca\,II\,8542;  
{\bf [b]} K\,I\,7699 with Zeeman splitting = 3000 G and q=1.3; {\bf [c]} Fe\,I\,5434 
with q=1.62; {\bf [d]} Na\,D$_2$ (only blue wing) with q=1.57. For comparison the 
photospheric (({\it dashes}) and penumbral (({\it dash-points}) profiles are shown.}
   \label{Fig1}
    \end{figure*}

%
%
                      
%

   \begin{figure}[h]     
   \hspace{0mm}\includegraphics[width=9.0cm]{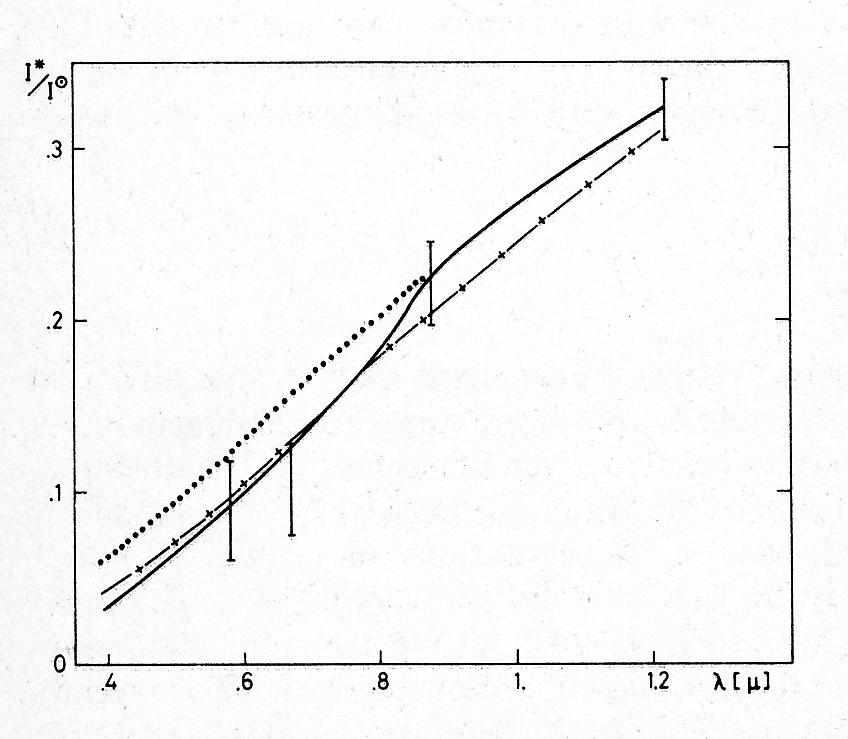}
   \caption{Wavelength dependence of the umbra-to-photosphere continuum intensity 
calculated for the models M3 (- x - x) and the new model M4 from this paper 
with $q(\lambda)=1.0$ (dots), resp., $q=q(\lambda)$ (full line). Bars give the range 
of observations by Ekmann \& Maltby (1974, cf. Stellmacher \& Wiehr, 1975, 
Fig.\,1).}
   \label{Fig2}
    \end{figure}

%
%

Figure 1 shows calculated profiles of Ca\,II\,8542, K\,I\,7699, NaD$_2$ and 
Fe\,I\,5434 (g\,=\,0) using different empirical models. It can be seen that the 
strong criterion of the unchanged Ca\,II lines is only reproduced by Zwaan's (1974) 
umbral model. Closer inspection shows that this agreement is essentially due to 
the high temperature gradient in the region $-0.6<log \tau_{0.5}<+0.3$, where 
the line wings are formed. On the other hand, it was shown by Stellmacher and Wiehr 
(1975) that Zwaan's (1974) model does not reproduce the wings of NaD$_2$ and Fe\,5434 
within the observational limits given by vanishing and, respectively, maximum 
correction for parasitic light. This also holds when considering our observations of 
the K\,7699 line wings (see Fig.\,1). The K line shows a similar temperature 
sensitivity as NaD$_2$, but it is much less influenced by parasitic light due 
to the almost two times smaller umbral continuum contrast at 7700\AA{} as compared 
to 5900\AA{}.

We tried to alter the temperature stratification of the Stellmacher and Wiehr (1975) 
model in order to obtain a better representation of the infrared Ca\,II lines without 
loosing the good fit of this model to the lines Fe\,5434, NaD$_2$ and also the newly 
included K\,7699 line. Our calculations show that those temperature stratifications which 
do fit the Ca\,II lines contradict the observations of either the profiles of Fe\,5434, 
NaD$_2$ and K\,7699 or of the relative umbra-to-photosphere continuum intensities 
(of $<0.1$ at 5900\AA{} and $<0.2$ at 7700\AA{}; see e.g. Zwaan, 1974). Preserving the 
temperature gradient required to fit the Ca\,II lines and shifting the $T(log \tau_{0.5})$ 
curve towards higher T (at constant $log \tau_{0.5})$, a good representation of Fe\,5434, 
NaD$_2$, and K\,7699 is achieved. A shift towards lower T, however, is required to fit 
the observed continuum contrasts. Any 'compromise' model between these two cases clearly 
yields Ca\,II lines and continuum contrasts significantly above, and Fe\,5434, NaD$_2$, 
K\,7699 profiles significantly below the range defined by uncorrected and maximum 
corrected observations. 
                      
%

   \begin{figure}[ht]     
   \hspace{0mm}\includegraphics[width=9.0cm]{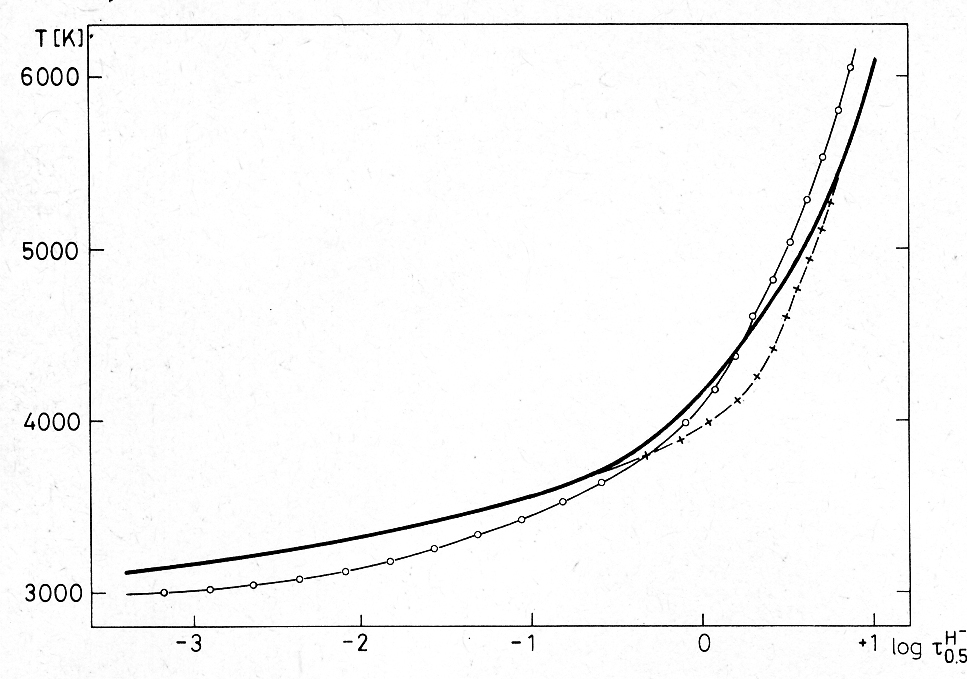}
   \caption{ Temperature versus optical depth $log \tau_{0.5}$ without opacity enhancement 
for the models by Zwaan (- o - o), our former model M3 (- x - x), and the improved model 
M4 from this paper (({\it full line}).}
   \label{Fig3}
    \end{figure}

%
%

%

   \begin{figure}[ht]     
   \hspace{0mm}\includegraphics[width=8.5cm]{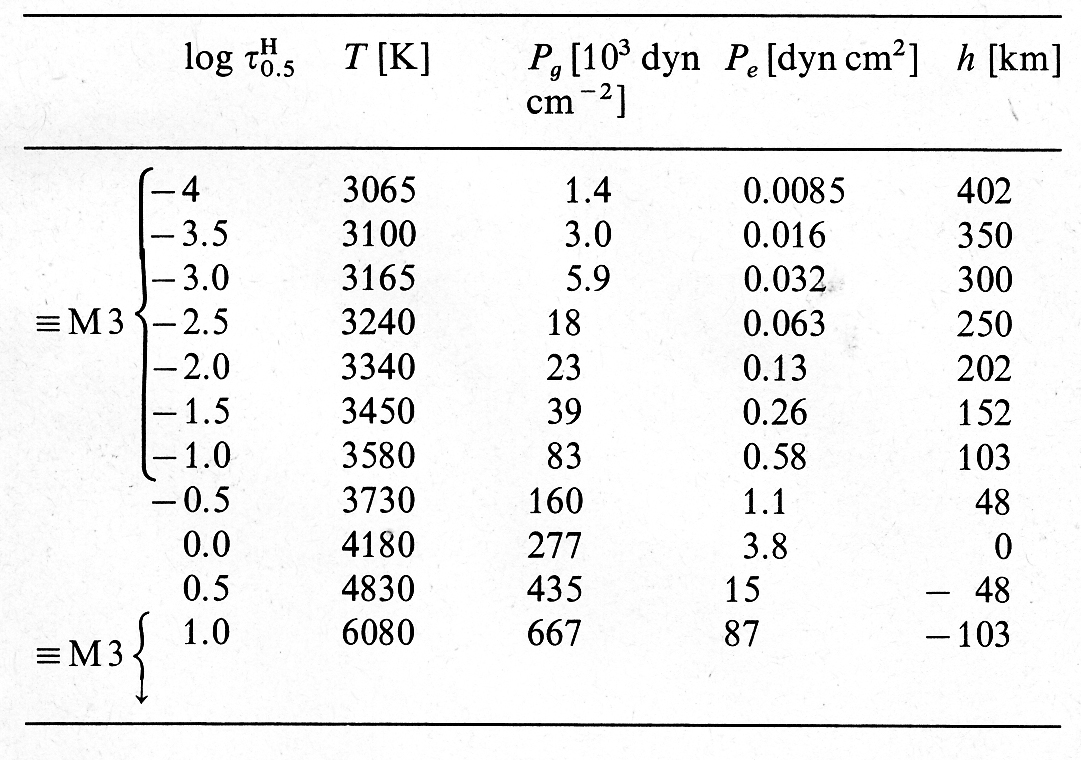}
   \caption{{\bf TABLE} with characteristic parameters of the final umbral model M4;
for $log \tau_{0.5}<-0.6$ and $log \tau_{0.5}>+0.6$ the preent model is identica with
our former model M3.}
   \label{Fig4}
    \end{figure}

%
%

\section{Discussion}

A reasonable fit to all observed lines and continuum data (see Fig.\,2) could be achieved 
when introducing the opacity enhancement $q(\lambda)$ for $\lambda<8500$\AA{} as suggested 
by Zwaan (1974). Several alterations of the $q(\lambda)$ dependence did not improve this fit.
The concept of an opacity enhancement for sunspot umbrae had been proposed since the wavelength 
variation of umbral continuum intensities at $0.5\mu<\lambda\le1.7\mu$ cannot be reproduced 
by a model even when assuming a maximum temperature gradient from a radiative equilibrium 
stratification. 

On the other hand, Stellmacher and Wiehr (1975) argue that the range of observed umbral 
continuum data and the uncertainties of the photospheric reference intensities still allow
a radiative equilibrium stratification without introducing an opacity enhancement. In that 
case, their model M3 gives the $0.5\mu$ intensity at its upper and the $1.7\mu$ intensity 
at its lower observational limits (Stellmacher and Wiehr, 1975, cf. Fig. 1). On the other 
hand, model M3 nicely describes the observed Fe\,5434 and NaD$_2$ line wings.

The consideration of observed infrared Ca\,II lines in addition that of Fe\,5434, NaD$_2$, 
K\,7699 and the corresponding continuum data gives further support to the existence of an 
enhanced opacity, which originally had been proposed by Zwaan (1974) for a good representation 
of umbral continuum contrasts alone. 

Our present observations do not allow to essentially refine the $q(\lambda)$ given by Zwaan. 
In particular, we can not verify whether $q(\lambda)$ further increases at $\lambda<5000$\AA{}, 
which might give some hint to the still unknown physical nature of the opacity enhancement. 
A possible explanation is line haze produced by atomic and molecular lines (Zwaan, 1974; 
Gaur et al., 1979) and/or unknown continuum absorbers (as e.g. proposed by Oinas, 1974, 
1977, for cool stars).

\subsection{Umbral model}

Our improved model 'M4', is given in Fig.\,3 together with our former model M3 and with 
Zwaan's (1974) model. The new model M4, listed in the table, differs from M3 only for
the layers $-0.6<log \tau{0.5}<+0.3$ in order to fit to the observed infrared Ca\,II 
lines discussed in this paper. For the outer layers $log \tau_{0.5}<-0.6$ M4 is identical 
with M3 to preserve the description of observed non-magnetic lines as discussed by 
Stellmacher and Wiehr (1970 and 1972). Also the deep layers $log \tau_{0.5}>0.6$ of M4 
agree with M3 in order to meet the radiative equilibrium stratification, which is required 
for $log \tau_{0.5}>0.8$ to fit the infrared C\,I lines discussed by Stellmacher and 
Wiehr (1975).  

The final umbral model M4 then optimally describes the whole range of continuum and line 
profile observations. It can thus be considered as an {\it empirical working model} with 
wide validity for dark umbral cores. Its temperature stratification is close to a scaled 
model of the undisturbed Sun with a scaling factor of $T_{eff}^{umbra}=4000$K and thus to 
a M0 rather than a K5 stellar atmosphere.

%
%

\begin{acknowledgements}
We thank H.\,Schleicher for helpful discussions and his LTE code. The Locarno 
observatory is operated by the German Science Foundation, DFG.   
\end{acknowledgements}

%
%

%
%

\end{document}